\documentclass[journal]{IEEEtran}

\makeatletter

\newcommand{\Rmnum}[1]{\expandafter\@slowromancap\romannumeral #1@}
\makeatother
\usepackage{graphicx,cite,epsfig,amssymb,amsmath,amsfonts,multirow}
\usepackage{mathrsfs}
\usepackage{float}
\usepackage{soul}
\usepackage{colortbl, color}
\usepackage{cases}
\usepackage[ruled,vlined,linesnumbered]{algorithm2e}
\usepackage{verbatim}

\usepackage{ulem}
\newcommand{\ls}[1]  
   {\dimen0=\fontdimen6\the=#1\dimen0
    \advance\lineskip.5\fontdimen5\the\lineskip-\dimen0
    \lineskiplimit=.9\lineskip
    \baselineskip=\lineskip
    \advance\baselineskip\dimen0
    \normallineskip\lineskip
    \normallineskiplimit\lineskiplimit
    \normalbaselineskip\baselineskip
    \ignorespaces
   }

\begin{document}
\bibliographystyle{ieeetr}

\title{Base Station ON-OFF Switching in 5G Wireless Networks: Approaches and Challenges}


\author{Mingjie~Feng,~\IEEEmembership{Student~Member,~IEEE},~Shiwen~Mao,~\IEEEmembership{Senior~Member,~IEEE}~and~Tao~Jiang,~\IEEEmembership{Senior~Member,~IEEE}%

\thanks{M. Feng and S. Mao are with the Department of Electrical and Computer Engineering, Auburn University, Auburn, AL 36849-5201 USA. Tao Jiang is with the School of Electronic Information and Communications, Huazhong University of Science and Technology, Wuhan, 430074, China. Email: mzf0022@auburn.edu, smao@ieee.org, tao.jiang@ieee.org.
}
\thanks{This work is supported in part by the
NSF under Grants CNS-1247955 and CNS-1320664, the Wireless Engineering Research and Education Center (WEREC) at Auburn University, and the National Science Foundation for Distinguished Young Scholars of China with Grant number 61325004.
}
}

\maketitle

\pagestyle{plain}\thispagestyle{plain}

\begin{abstract}
To achieve the expected 1000x data rates under the exponential growth of traffic demand, a large number of base stations (BS) or access points (AP) will be deployed in the fifth generation (5G) wireless systems, to support high data rate services and to provide seamless coverage. Although such BSs are expected to be small-scale with lower power, the aggregated energy consumption of all BSs would be remarkable, resulting in increased environmental and economic concerns. In existing cellular networks, turning off the under-utilized BSs is an efficient approach to conserve energy while preserving the quality of service (QoS) of mobile users. However, in 5G systems with new physical layer techniques and the highly heterogeneous network architecture, new challenges arise in the design of BS ON-OFF switching strategies. In this article, we begin with a discussion on the inherent technical challenges of BS ON-OFF switching. We then provide a comprehensive review of recent advances on switching mechanisms in different application scenarios. Finally, we present open research problems and conclude the paper.
\end{abstract}


\section{Introduction}

With the growing popularity and versatility of mobile devices, the age of mobile Internet has come. According to a report from comScore Inc., mobile devices accounted for 55$\%$ of Internet usage of the United States in January 2014, which was the first time that mobile apps (applications) overtake personal computer (PC) for Internet usage in the US. As a result, the requirements for data-intensive wireless services are boosted at an unprecedented rate, posing great pressure on current wireless systems. To satisfy such demand, the fifth generation (5G) wireless system is under intensive development,
and it is expected to provide ubiquitous Internet access with 1000x data rate. Compared to previous generations of mobile communication systems, the 5G system not only involves innovations in the physical layer techniques, but also introduces new network architectures and application scenarios. In particular, a major trend in 5G networks is the deployment of a large number of small-scale BSs or APs, also known as {\em network densification}. For example, in mmWave network, small cell network, distributed antennas system, femtocaching-enabled network, BSs are deployed close to users to combat propagation loss, to improve signal to noise ratio (SNR), and to reduce service delay. However, these benefits come at a price, the massive BS deployment significantly increases the total energy consumption of wireless systems. For a typical LTE microcell with cell size of 100 m and bandwidth 5 MHz, the power consumption is ranged from 25 watts to 40 watts depending on the traffic load. To achieve the coverage of a 1500 m macrocell, more than 200 microcells need to be deployed, the aggregated power of microcells can be more than 900 watts, which is comparable to a typical LTE macrocell BS with 1500 m coverage. The increased energy consumption not only increases the cost of wireless operators, but also generates more greenhouse gas emissions. Thus, energy saving has become an important design objective of wireless systems in recent years. Meanwhile, energy saving needs to be achieved without sacrificing the quality of service (QoS) of users. As the 5G system is expected to provide 1000x data rates, energy efficiency (EE), typically measured by bits/Joule, also needs to be increased by 1000 times if the total energy consumption maintains at its original level.

BS ON-OFF switching (also known as BS sleep control) has been considered as an efficient approach for both energy saving and EE improvement. As the traffic pattern fluctuates over both time and space, under-utilized BSs can be dynamically turned off to save energy~\cite{Niu11}. From 2009, China Mobile began to apply BS sleep control and the estimated reduction of energy consumption is 36 million kWh per year. Due to such great potential, considerable efforts have been devoted to the design of BS ON-OFF switching strategies in different network scenarios. However, as the 5G system is an integration of different techniques with a highly heterogeneous network architecture~\cite{Andrews14}, the design of BS ON-OFF switching faces special challenges in 5G systems, which can be summarized as follows.
\begin{itemize}
\item {\em Interoperability with New Technologies}: With new technologies in 5G systems, additional constraints and impacts emerge. It is necessary to adjust the BS ON-OFF switching strategy accordingly. For example, when device to device (D2D) communication is available for user equipments (UE), a UE can still send/receive data to/from a BS with the help of another UE even when it is not within the coverage of the BS. In this case, some BSs can be turned off without worrying about the resulting coverage holes. As another example, when 5G systems operate on unlicensed bands that are currently used by other systems (e.g., WiFi), the BS ON-OFF pattern may be accommodated to maintain the QoS of WiFi users.
\item {\em Applications in New Network Architectures}: In the new 5G network architecture, the functionalities and properties of BSs are significantly changed. For instance, in cloud radio access network (C-RAN), users are connected to remote radio units (RRU) while the data processing is performed by centralized baseband units (BBU). Since a BBU operates with much higher power than an RRU, turning off the under-utilized BBUs is desirable for energy saving. However, when a BBU is switched off, all RRUs linked to the BBU cannot provide service to users. To prevent outage of users, either these RRUs connect to a new BBU or the users are handover to other available RRUs, resulting in a complicated scheduling problem. As another example, when a BS in a millimeter wave (mmWave) network is turned off, its coverage can hardly be compensated by neighboring BSs due to high propagation loss and blockage by obstacles. Hence, the QoS of users cannot be guaranteed with the existing BS ON-OFF scheduling and new approaches are required to address this issue.
\item {\em Large Number of BSs}: To satisfy the high aggregated data rate requirements in hotspots, the BSs may be densely deployed with a certain level of redundancy. Thus, turning off a BS increases the traffic loads of multiple nearby BSs. The ON-OFF states and QoS provision between different BSs could be highly interdependent. With lots of low-power BSs, a significant amount of BSs need to be switched off to achieve energy saving. This results in a combinatorial problem with large number of variables, which is generally difficult to solve with standard techniques. In addition, BS ON-OFF switching also impacts the interference pattern if these BSs operate on the same spectrum band. In case coordinated multipoint (CoMP) transmission is available, each user can be served by multiple BSs, the problem would be more complicated. The large number of BSs also cause scalability issues, which is a key factor that affect the feasibility and effectiveness of scheduling algorithms.
\end{itemize}

To harvest the benefits of BS ON-OFF switching under these challenges, it is necessary to investigate the technical aspects of switching mechanisms and analyze its challenges in the new 5G system context, where different techniques and network architectures are integrated. This article aims to identify the key challenges on BS ON-OFF switching and provide insights to its applications in 5G systems. We first analyze the technical aspects and challenges of ON-OFF switching operation, followed by a review of the recent advances in different wireless systems. Then, we discuss open research problems in some emerging application scenarios and present potential solutions.

\section{Technical Aspects and Challenges of BS ON-OFF Switching \label{sec:tech}}

\begin{table*}
\vspace{0.5in} \centering \caption{Different Aspects of System Model Regarding BS ON-OFF Switching} \label{tab1}
\begin{tabular}{|c|c|c|c|c|}
\hline
\multirow{2}*{\bf{}} & \multirow{2}*{\bf{BS Type}} & \multirow{2}*{\bf{BS Energy/Power Model}} & \multirow{2}*{\bf{Traffic Type}} & \multirow{2}*{\bf{Timescale}}   \\ & & & &  \\
\hline
\multirow{2}*{\cite{Son11}} & \multirow{2}*{Regular BS} & ON: Fixed + load-proportional; & PPP based traffic arrival; & Slow: BS ON/OFF switching;  \\ & & OFF: 0. & log normally distributed file size & Fast: user association.  \\
\hline
\multirow{2}*{\cite{Oh13}} & \multirow{2}*{Regular BS}  & \multirow{2}*{Fixed value}  & Poisson arrival UEs, & Slow: duration in objective function;  \\ & & & exponential file size. & Fast: update of traffic pattern. \\
\hline
\multirow{2}*{\cite{Guo16}} & \multirow{2}*{Regular BS} & Active: idle power + load-dependent; & Poisson arrival tasks; & Hysteresis sleep time  \\ & & Sleep: sleep power + detection power. & general distribution service time. &  and wake up period \\
\hline
\cite{Zhang15} & Small cell BS & Fixed value & PPP based user distribution & Traffic pattern update period \\
\hline
\multirow{2}*{\cite{Liu16}} & \multirow{2}*{Small cell BS} & Four modes: ON (100\%), standby (50\%), & \multirow{2}*{PPP based user distribution} & Slow: BS ON/OFF switching \\ & & sleep (15\%), and OFF (0\%). & & Fast: time required to wake up  \\
\hline
\multirow{2}*{\cite{Cai16}} & Macrocell BS and & MBS: Fixed + load-dependent, & Uniform and non-uniform  & \multirow{2}*{BS ON/OFF switching}  \\ & Small cell BS & SBS: fixed for ON and OFF. & PPP based user distribution &  \\
\hline
\multirow{2}*{\cite{Feng16}} & \multirow{2}*{Small cell BS} & \multirow{2}*{Fixed value}  & Uniform and non-uniform  & Slow: BS ON/OFF switching  \\ & & & random user distribution & Fast: user association \\
\hline
\multirow{2}*{\cite{Wang16}} & \multirow{2}*{BBU and fronthaul} & Proportional to the numbers of & A given traffic distribution & Period for network planning   \\ &  &  active BBUs and fronthaul links. & on daily basis & Period for traffic engineering  \\
\hline
\multirow{2}*{\cite{Zhang14}} & \multirow{2}*{Multi-antenna BS} & Four parts: circuit + backhauling & A certain amount of & Instantaneous, average CSI,  \\ &  &  + transmission + processing. & users to be served & and BS ON/OFF switching.  \\
\hline
\multirow{3}*{\cite{Gong14}} & \multirow{3}*{BS with EH} & Active: fixed + $\frac{\rm{RF~power}}{\rm{Efficiency}}$. & Poisson arrival UEs; & \multirow{3}*{Period for metric evaluation}  \\ &  &  Opportunistic sleep: a fraction of active. & given service rate; &   \\  & & Deep sleep: 0. & user data rate requirements. &  \\
\hline
\multirow{2}*{\cite{Zhang16}} & \multirow{2}*{BS with EH} & Active: Fixed + $\frac{\rm{RF~power}}{\rm{Efficiency}}$. & \multirow{2}*{PPP based user distribution} & Large: user density, harvest rate;  \\ &  &  Sleep: 0. & & Small: user location, battery level.  \\
\hline
\multirow{2}*{\cite{Niyato12}} & BS powered by & \multirow{2}*{On grid + renewable} & 5 scenarios with & Long-term: 1 day;  \\ & Smart grid  &   & different arrival rates & Short term: 1 hour. \\
\hline
\cite{Antonopoulos15} & Regular BS & Fixed + load-dependent & Not specified & Note specified  \\
\hline
\end{tabular}
\end{table*}

\subsection{Energy Consumption Model}

The BS energy consumption model has evolved from simple and approximated models to more sophisticated ones, depending on the BS type and application scenario. A BS consumes a certain amount of energy to maintain its normal operation, such as energy for circuits, cooling system, etc. Since such static parts constitute a dominant proportion of the total BS energy consumption, binary models are used to approximate the energy consumption under ON-OFF states~\cite{Oh13,Zhang15,Feng16}. To capture the impact of traffic load, the dynamic load-dependent part was also considered. For example, the dynamic part can be proportional to the number of serving users~\cite{Son11} or to the number of active antennas~\cite{Zhang14}.

The BS power under the sleep mode is regarded as zero in some works, since it is small compared to the power when the BS is fully active. However, to ensure that
a BS can be activated, the BS should not be completely powered off; it may still consume a certain amount of power, such as detection power~\cite{Guo16}.
Such power is not negligible, especially for small-scale BSs without on-site cooling systems. Thus, the power under sleep mode is also studied in recent models,
and the sleep mode is further classified into different levels of sleep. In~\cite{Gong14}, the concept of opportunistic sleep is proposed, in which a BS only
sleeps in certain time periods to improve system reliability. In~\cite{Liu16}, the BS states are further classified into four types, including ON, standby, sleep, and OFF,
with power consumption ratios given as 100 \%, 50\%, 15\%, and 0\%, respectively.

\subsection{Traffic Model and Traffic-Aware Scheduling}

The traffic models used in prior works are summarized in Table~\ref{tab1}. Among these models, the key factors include user arrival rate, user distribution over space, file/task size, and service rate. Based on these parameters, the stochastic geometry framework was widely used to analyze the theoretical system performance. However, the traffic demands from users are heterogeneous in nature, which cannot be characterized by a single traffic model. For example, some demands are delay-sensitive (e.g., a phone call), some are rate-sensitive (e.g., a file download), some are both delay and rate sensitive (e.g., online gaming and video conferencing), while some are neither (e.g., information gathered from sensor network). Considering that BSs can work in different modes with different service provisioning capabilities, one can adjust the operating modes of BSs according to the traffic type and pattern to further enhance the system performance.

We use a simple and highly abstracted indicative example to show the potential of traffic-aware scheduling. Consider a given area with multiple BSs, each BS has two sleep states, namely {\em standby} and {\em deep sleep}. The standby state consumes higher power but is faster to wake up
compared to the deep sleep state. We assume that BSs in both states wake up periodically and it takes a certain amount of energy to wake up. Without loss of generality, we consider an example of traffic pattern with half of the traffic demands are delay-sensitive and the other half are rate-sensitive. To fully utilize the advantages of BSs in both states for both energy saving and delay reduction, it is obvious that half of the BSs should be in standby states to serve the delay-sensitive demands while the other half in deep sleep to serve the rate-sensitive demands. We call such a schedule {\em adaptive partial scheme} and compare it with other two schemes with only one BS sleep state.

\begin{figure}[!t]
 \begin{center}
   \includegraphics[width=3.0in]{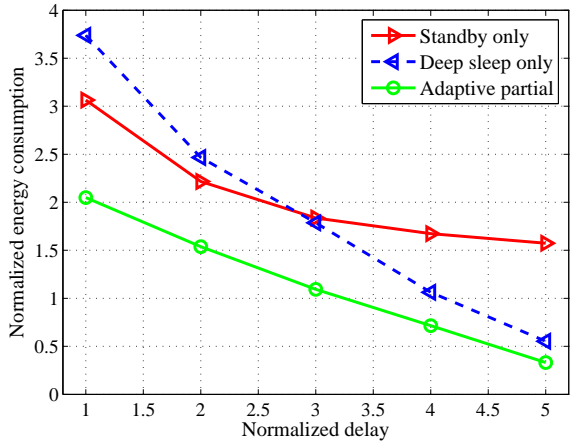}
 \end{center}
\caption{Energy-delay tradeoff under different sleep modes with heterogeneous traffic requirements.}
\label{fig1}
\end{figure}

In Fig.~\ref{fig1}, we obtain different energy-delay pairs by changing the wake up intervals. For example, a standby BS reduces its energy consumption by prolonging its duration of sleep; a deep sleep BS reduces its service delay by waking up more frequently and sleeping for shorter durations. It can be seen in Fig.~\ref{fig1} that the adaptive scheme outperforms both benchmark schemes, indicating that a considerable performance gain can be achieved by adjusting the operation states of BSs according to traffic type.

\subsection{Practical and Implementation Concerns}

\subsubsection{Timescale of Operation}

Since the ON-OFF states directly determine the QoS of users, BS ON-OFF switching is always coupled with other design issues, such as user association and traffic offloading. As discussed, BS ON-OFF switching takes both time and energy; it is thus infeasible to perform ON-OFF switching frequently. However, the system states are usually updated at a faster pace, and some technical approaches have to be executed more frequently. For example, due to user mobility, user association is updated more frequently than BS ON-OFF switching. In addition, as the performance of a wireless system largely depends on the channel condition, which is rapidly changing, it is necessary to consider time-averaged channel state information (CSI) instead of instantaneous CSI. The timescale issue of different scenarios are also summarized in Table~\ref{tab1}.

\subsubsection{How to Acquire System Information and Wake Up When Sleeping?}

When a BS is turned off, the normal transmission between the BS and UEs is suspended. Hence, the information of nearby UEs, such as CSI and traffic load, cannot be acquired by the BS from the uplink signals. To guarantee the effectiveness of scheduling, the BSs need to be aware of the environment even when they are in the sleep mode, so that they can be activated on a timely manner. In~\cite{Oh13}, waking up a BS is performed with the assistance of neighboring BSs. When a BS is switched off, the neighboring BSs record their own system loads, and use such information as the criterion for mode switching. This way, the neighboring BSs know when and under what conditions a BS should be turned on, and inform the BS to do so.

Alternatively, a BS can wake up itself periodically to detect the environment. It returns to sleep when the criterion for switching on is not satisfied~\cite{Liu16,Gong14,Guo16}. Under this model, the wake up interval, which determines the wake up frequency, is a key design factor. We show that the wake up interval can be optimized under different sleep modes and traffic patterns with an illustrative example below.

Consider a heterogeneous network (HetNet) with one macrocell BS (MBS) and multiple small cell BSs (SBS). The MBS is always acvive to guarantee coverage, while the SBSs can be dynamically switched on or off. The SBSs can operate in two modes, the standby mode and the deep sleep mode.
We consider both high mobility and low mobility scenarios. In the high mobility scenario, the number of users in a small cell has a larger variation compared to that
in the low mobility scenario.
The system configuration from~\cite{Feng16} is used in the simulation. We compare the EE performance under different wake up intervals.

\begin{figure}[!t]
 \begin{center}
   \includegraphics[width=3.0in]{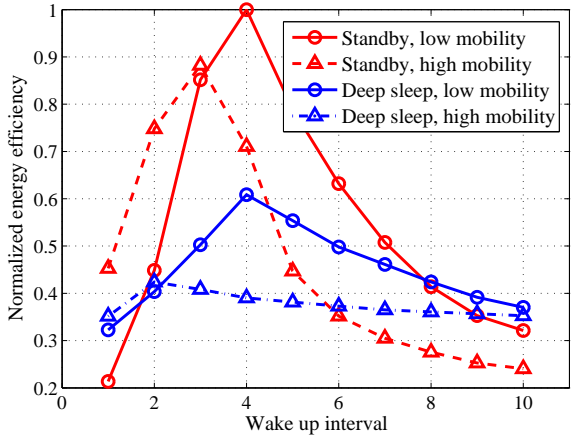}
 \end{center}
\caption{EE versus wake up interval under different sleep modes and levels of mobility.}
\label{fig2}
\end{figure}

As shown in Fig.~\ref{fig2}, a proper value of wake up interval can maximize the EE in all the scenarios. The standby mode outperforms the deep sleep mode when the wake up interval is small, since it can closely capture the dynamics of traffic load. When the wake up interval becomes large, the BSs cannot obtain the load information on time. The standby mode losses its advantage while it consumes more energy, resulting in lower EE compared to deep sleep. As expected,
a short wake up interval is preferred in the high mobility scenario.

As another option, the system information can also be obtained with auxiliary devices when a BS is in the sleep mode. In~\cite{Guo16}, a counting element is used to count the number of accumulated tasks during a sleep period, and a BS wakes up when the number of tasks exceeds a threshold. To address the problem of obtaining CSI of nearby UEs, a possible approach is to deploy a radio receiver at a BS to capture the pilot or uplink UE signals, and forward the obtained CSI to the network controller. Then, the network controller can activate the BS when necessary, e.g., when a large number of users with good channel conditions are nearby.

\subsection{Tradeoff with Other Performance Metrics}

The energy savings achieved by switching off BSs come at a price. From the perspective of users, some users have to handover to another BS and receive a degraded QoS. Besides, the BSs in the sleep mode may not be turned on timely, resulting in extra delay perceived by users. To characterize the tradeoff between energy saving and user QoS, a common approach is to add QoS requirements into the constraints of problem formulation~\cite{Guo16,Zhang15}. Another approach is to consider the actual relationship between energy consumption and system performance. EE, defined as bits/Joule, is a common performance metric to balance the tradeoff between data rate and power consumption~\cite{Liu16,Feng16,Zhang14}. In~\cite{Son11}, a cost function is defined for a balance between energy consumption and delay, which is defined as a weighted sum of energy consumption and delay.

When a BS is turned off, the traffic load of neighboring BSs would increase, which potentially degrades the QoS of users served by these BSs. Thus, it is also necessary to consider the tradeoff among different BSs. In a two-tier HetNet with a large number of SBSs turned off, the MBS may be overloaded, resulting in delay or even congestion. In~\cite{Cai16}, a constraint on the number of users served by the MBS is enforced to guarantee the QoS of users served by MBS.

\subsection{Low Complexity Algorithms}

As the BS ON-OFF scheduling is combinatorial in nature, it is generally NP-hard and cannot be solved with standard techniques. It is more challenging when BS ON-OFF is jointly scheduled for trade-offs among multiple performance metrics,
as the formulated problem may be mixed integer programming with multiple sets of variables. To decouple BS ON-OFF switching with other tasks, it is natural to decompose the original problem into several subproblems and iteratively solve them~\cite{Son11,Feng16,Gong14}. With regard to BS ON-OFF switching, step-by-step greedy algorithms are developed~\cite{Son11,Oh13,Cai16,Zhang14,Zhang16}. By evaluating the system performance gains under different operations, the BS that brings the largest gain is selected in each step. The optimal ON/OFF states can also be obtained by fully utilize the special properties of the problem~\cite{Guo16,Zhang15}. In~\cite{Zhang15}, with a comprehensive performance analysis, the optimal BS ON-OFF states are found to correspond to the case that equalities hold for outage constraints.

Another possible approach for low complexity solution is to transform the original problem into an amiable form that is solvable. In~\cite{Liu16}, to deal with the non-convex objective function, a lower bound that is quasi-convex is derived, which serves as the new objective function. The solution obtained by maximizing the lower bound is shown to be near-optimal. Designing distributed schemes is another way for low complexity solutions. In~\cite{Feng16}, a bidding game between users and BSs is formulated and the initial outcome of the game determines a user association strategy. When user associations are determined, each BS makes its own decision on ON/OFF switching depending on whether it is profitable to serve these users. As a result, BSs with low traffic loads would choose to turn off. The system EE can thus be improved. Table~\ref{tab2} provides a summary of different problem formulations and solution algorithms presented in the literature.

\begin{table*}
\vspace{0.5in} \centering \caption{Problem Formulations and Solution Algorithms} \label{tab2}
\begin{tabular}{|c|c|c|c|c|}
\hline
\multirow{2}*{\bf{}} & \multirow{2}*{\bf{Objective}} & \multirow{2}*{\bf{Constraint}}  & \multirow{2}*{\bf{Approach}} & \multirow{2}*{\bf{Solution}} \\ & & & & \\
\hline
\multirow{2}*{\cite{Son11}} & Minimize weighted sum & \multirow{2}*{N.A.} & BS ON-OFF switching & Decompose into two subproblems; \\ & of energy and delay &   & and user association. & Greedy ON-OFF scheduling.\\
\hline
\multirow{2}*{\cite{Oh13}} & Minimize energy over  & \multirow{2}*{BS traffic load} & \multirow{2}*{BS ON-OFF switching} & Step-by-step process with evaluation on  \\ & a period &  &  & the {\em network impact} of turning ON/OFF.  \\
\hline
\multirow{2}*{\cite{Guo16}} & \multirow{2}*{Minimize average power} & \multirow{2}*{Average delay} & Optimize hysteresis sleep & Simple bisection search based on \\ & &  & time and wake up period &  exploiting the special structure  \\
\hline
\multirow{2}*{\cite{Zhang15}} & Minimize the ratio of & \multirow{2}*{UE Outage probability} & BS ON-OFF scheduling & Solutions obtained when equalities  \\ & turned OFF small cells & & and spectrum allocation & hold for outage constraints \\
\hline
\multirow{2}*{\cite{Liu16}} & \multirow{2}*{Maximize EE} & Coverage probability & \multirow{2}*{BS sleep mode design} & Maximize a quasi-convex lower bound; \\ & & and delay & & An iterative scheme for solution. \\
\hline
\multirow{2}*{\cite{Cai16}} & \multirow{2}*{Minimize total BS power} & \multirow{2}*{Macrocell BS power}  & \multirow{2}*{Active/sleeping schedule} & Step-by-step process by evaluating \\ &  &  &  & the gain/loss of operation \\
\hline
\multirow{2}*{\cite{Feng16}} & \multirow{2}*{Maximize EE} & \multirow{2}*{BS traffic load}  & BS ON-OFF switching  & Decompose into two subproblems;  \\ & & & and user association & Centralized and distributed solution.  \\
\hline
\multirow{2}*{\cite{Wang16}} & Minimize the number & Bandwidth; Cell and user & Time-wavelength allocation; & Load balancing for overloaded BBU;   \\ & of active BBUs &  processing capability. & User-fronthaul-BBU mapping. & Activate a BBU if still overloaded. \\
\hline
\multirow{2}*{\cite{Zhang14}} & \multirow{2}*{Maximize EE} & Data rate, power, & BS and antenna switching; & Turn off the antenna/BS such that\\ &  &  and precoding & Power allocation. &   total power is minimized afterwards.  \\
\hline
\multirow{3}*{\cite{Gong14}} & \multirow{3}*{Minimize on-grid power} & \multirow{3}*{Blocking probability} & BS ON-OFF switching; & Two-stage dynamic programming;  \\ &  &   & Resource allocation; & 1st: BS ON-OFF;  \\  & &  & Renewable energy allocation. & 2nd: resource allocation.\\
\hline
\multirow{2}*{\cite{Zhang16}} & \multirow{2}*{Minimize on-grid power} & Outage probability and & BS ON-OFF switching & SBSs with
positive power saving  \\ &  &  resource availability & and offloading strategy &  gain are activated first.  \\
\hline
\multirow{2}*{\cite{Niyato12}} & \multirow{2}*{Minimize power cost} & Battery storage; Supply & Dynamic power usage  & Formulate and solve  \\ &  &  and consumption balance & of different sources &  a stochastic programming.  \\
\hline
\multirow{2}*{\cite{Antonopoulos15}} & Each operator minimizes & Rewards and penalties & Provide incentive for infrastructure & Analyze the outcome  \\ & its cost & of the game & sharing by formulating a game & of the game \\
\hline
\end{tabular}
\end{table*}

\section{Recent Advances in Emerging Wireless Networks \label{sec:app}}

\subsection{Renewable Energy Empowered Network}

The use of renewable energy can minimize the on-grid energy consumption as well as reduce the operating expenditure. This can be realized by equipping BSs with energy harvesting (EH) devices that transform natural energy (e.g., solar and wind energy) to electricity to power the BSs. However, as the generation of renewable energy depends on uncontrollable environmental conditions, the supply of renewable energy could be highly unstable. In addition, it is very likely that such supply does not match the fluctuating traffic demands in both temporal and spatial domains. Thus, the renewable energy should be properly stored and allocated for efficient utilization.

In~\cite{Gong14}, with statistic information of both renewable energy and traffic pattern, renewable energy allocation, BS ON-OFF scheduling, and resource allocation are jointly considered to minimize the on-grid energy consumption, subject to average blocking probability of users. The original problem is transformed to an unconstrained problem, and a two-stage dynamic programming algorithm with low complexity is proposed to obtain a near-optimal solution. The case of HetNet is considered in~\cite{Zhang16}, where the SBSs are equipped with EH devices. The switching mechanism of SBSs and traffic loading from MBS are considered to minimize the on-grid power while satisfying constraints on outage probability and resource availability. Based on theoretical analysis on outage probability, a two stage SBS activation scheme is proposed. In the first stage, the energy saving gain obtained by offloading MBS traffic to each SBS is obtained. In the second stage, SBSs with positive energy saving gain are activated first.

\subsection{Massive MIMO HetNet}

A massive MIMO HetNet employs a large number of antennas at the MBS~\cite{Feng16}. The channel estimation overhead is a major concern
due to the large dimension of channel matrix. When the traffic load of MBS is increased, more symbols must be used as pilots in each frame. As a result, a less proportion of symbols can be used for data transmission and the average throughput of users will be reduced. Thus, in a massive MIMO HetNet, the switching mechanism needs to consider the tradeoff between the traffic loads of MBS and SBSs. In~\cite{Feng16}, joint BS ON-OFF scheduling and user association are considered to maximize the EE of a massive MIMO HetNet, subject to traffic load constraints of all BSs. Due to the convexity and a special property, the optimal user association strategy under a given BS set can be achieved using a series of lagrangian dual methods. Then, ON-OFF states can be optimized with a subgradient method using the optimal lagrangian multipliers derived in the user association subproblem. A distributed scheme based on user bidding is also proposed, which has been discussed in Section~\ref{sec:tech}-E.

\subsection{Cloud Radio Access Network}

In C-RAN, the functionality of a BS is separated: data processing is performed by centralized BBUs and wireless signal transmission/reception is performed by geographically distributed RRUs. The ON-OFF switching of RRUs is relatively simple and some existing methods based on traditional BS model can be directly applied. In addition, since the RRUs only serve as transmitters/receivers with low power consumption, the energy saving gained from turning off RRUs alone would be limited. Thus, it is highly appealing to investigate the ON-OFF switching for BBUs. However, turning off a BBU impacts all its serving RRUs and users connected to the RRUs. Hence, the ON-OFF states of BBUs are coupled with BBU-RRU and RRU-user mappings, resulting in a challenging scheduling problem. Such problem is termed {\em virtual BS formulation} in~\cite{Wang16}, and a joint time-wavelength allocation for optical fronthaul, user-RRU-BBU mapping, and BBU ON-OFF scheduling scheme is proposed to minimize the number of active BBUs. The idea of solution algorithm is to find the overload BBUs, and then apply load balancing
for the overloaded BBUs. If any BBU is still overloaded, a new BBU has to be activated. This process can guarantee that all constraints are satisfied, and the number of active BBUs is kept as small as possible.

\subsection{Cooperative Communication}

Coordinated multipoint (CoMP) transmission/reception, which allows a user to be served by multiple cooperating BSs, is an effective approach to enhance spectral efficiency and link reliability at the price of increased overhead. Under this context, detailed physical layer signal analysis is required to study the impact of BS ON-OFF switching. In~\cite{Zhang14}, power allocation and ON-OFF switching for both BS and antenna are considered to maximize the EE of a CoMP system. A low complexity iterative greedy approach is proposed, the idea is to switch the antenna/BS that reduces most power consumption until any constraint is violated.

\subsection{Smart Grid Empowered Wireless System}

The smart grid is a new paradigm of power systems which enables highly efficient use of energy, especially renewable energy. Thus, using smart grid to power wireless systems is a promising approach to reduce the operation cost and greenhouse gas emission. For example, the BS of a wireless system can be powered by a combination of renewable power source and electrical grid. However, this approach is challenged by a set uncertainties, including renewable power generation, power price, and wireless traffic load. In~\cite{Niyato12}, an adaptive demand-side power management scheme is proposed to make intelligent decisions on the power usage between renewable power and on-grid power. Such a problem is formulated as a stochastic programming with the objective of minimizing the cost of power consumption, and the optimal solution is derived by solving an equivalent linear programming problem.

\subsection{Infrastructure Sharing Among Different Operators}

While most existing works consider BS ON-OFF switching from the perspective of a single wireless operator, infrastructure sharing among different operators has the potential to achieve better system performance. As the BSs of multiple operators coexist in a cell, an operator can switch off its BS and migrate
its traffic to the active BSs of another operator covering the same area. To enable such a process, it is necessary to design an incentive mechanism to motivate the cooperation of operators. In~\cite{Antonopoulos15}, a distributed game of multiple operators is formulated to provide incentive for infrastructure sharing, and each operator can make rational decision on its switching strategy.

\section{Challenges and Open Problems in 5G \label{sec:future}}

\subsection{5G Application Scenarios}

\subsubsection{D2D Communications}

With the help of D2D communications,
a BS can be turned off even if this leads to coverage holes, resulting in more energy saving. However, this comes at a price of increased energy consumption at UEs. Thus, it is necessary to motivate the participation of UEs and efficiently use the limited power of UEs. BS ON-OFF switching should also be jointly considered with other design issues, such as D2D path selection and resource allocation. For example, suppose UE 1 relays the signal of UE 2 to a BS. To avoid congestion, the data rate between the BS and UE 1 should be no less than the data rate between UE 1 and UE 2 plus the actual data rate of UE 1. With such constraints, BS ON-OFF switching depends on the availability of UEs, link scheduling, and resource allocation. All these factors should be considered in future research.

\subsubsection{MmWave Networks}

MmWave communications is also a key technology for 5G wireless.
Due the large propagation loss and vulnerability to blockage, the signal transmission in a mmWave network can only rely on line of sight transmission or reflection. As a result, when a BS is turned off, its coverage may not be easily compensated by other BSs. A possible approach to deal with this challenge is to optimize the deployment pattern of BSs. With proper setting of BS locations and well designed beamforming schemes, the coverage holes caused by turning off BSs can be minimized. In case when direction of arrival (DoA) estimation is feasible, the instantaneous positions of UEs can be obtained by each BS. Then, a BS can know whether a UE can be served by another BS based on the layout of environment, and make decision on its ON-OFF state.

\begin{figure}[!t]
 \begin{center}
   \includegraphics[width=3.3in]{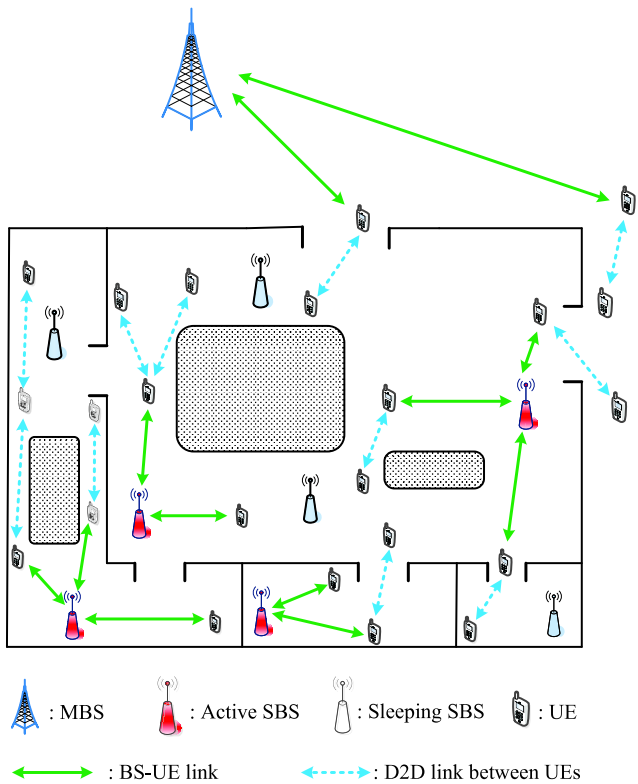}
 \end{center}
\caption{An example of BS ON-OFF scheduling in a D2D assisted mmWave network.}
\label{fig3}
\end{figure}

D2D communication is an effective means to combat blockage of signals in mmWave network. As shown in Fig.~\ref{fig3}, D2D communication can significantly enhance the coverage of both indoor and outdoor UEs. Apart from the issues mentioned before, some inherent challenges of mmWave network also need to be considered, such as neighbor discovery and efficient beamforming. Moreover, the UEs may not be able to perform ``pseudo-wired'' beamforming, and their transmission signals may still have a large beamwidth.
Interference management is necessary to guarantee the link qualities.

\subsubsection{Heterogeneous Networks with Wireless Backhaul}

\begin{figure}[!t]
 \begin{center}
   \includegraphics[width=3.6in]{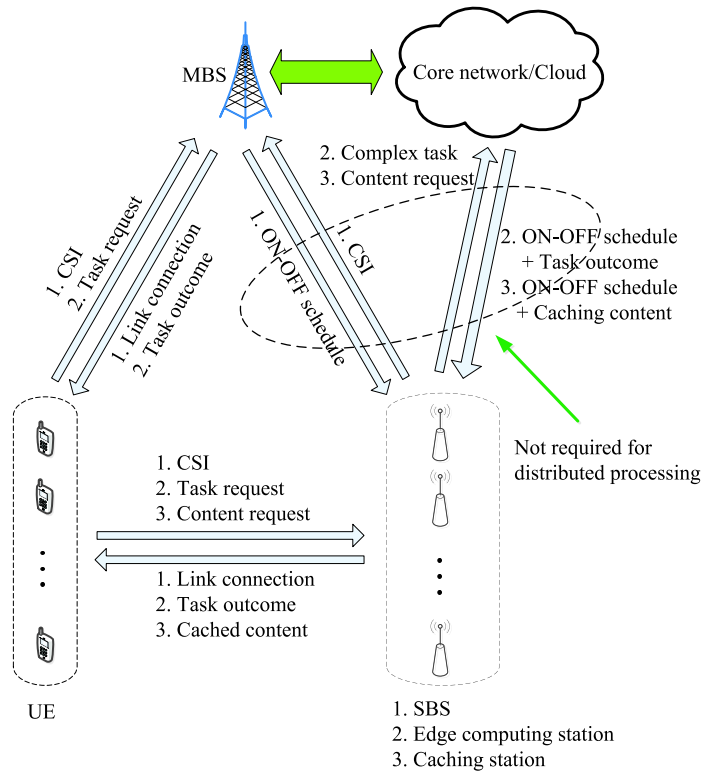}
 \end{center}
\caption{Network architecture and information exchange in $3$ heterogeneous application scenarios.}
\label{fig4}
\end{figure}

Wireless backhaul (WB) between MBS and SBSs in a HetNet has gained growing attention due to its easy implementation and relatively low cost. As an inherent constraint, the data rate of WB between the MBS and an SBS should be no less than the aggregated data rate of all small cell UEs (SUE) served by the SBS. Hence, without proper configuration, WB may become a bottleneck that limits the system performance. When switching off an SBS, the aggregated data rate requirement of neighboring SBSs would be increased, putting pressure on their WBs. Thus, guaranteeing the data rates of WBs is a key concern when SBS ON-OFF scheduling is enabled. From the perspective of MBS, the WB can be regarded as a macrocell UE (MUE) to be served. Hence, the data rates of WBs are also related to the traffic load and scheduling strategy of the MBS, and this is an important factor for system design. The network architecture of a HetNet is shown in the {\em 1st} case of Fig.~\ref{fig4}, it can seen that the wireless backhaul also needs to transmit the information exchange between MBS and SBSs. Thus, guaranteeiung the effectiveness and timeliness of scheduling in another design factor for a HetNet with wireless backhaul.

As massive MIMO is a key technique for 5G, we consider a massive MIMO HetNet as an example, in which the MBS is equipped with a large number of antennas and the SBSs can be turned off when their load is low. The WBs and MUEs are put into several beamforming groups and the channel estimation overhead is proportional to the number of active WBs and MUEs. When an SBS is turned off, the UEs originally served by the SBS may handover to the MBS or neighboring SBSs depending on the CSI. If the UEs handover to the MBS, the average throughput of active WB and MUE would decrease, as analyzed in~\cite{Feng16}. If the UEs handover to neighboring SBSs, the WBs of these SBSs need to be allocated with more spectrum resource to maintain the average throughput of UEs served by these SBSs. All these tradeoffs need to be fully balanced for the design of an energy efficient massive MIMO HetNet.

\subsubsection{Mobile Edge Computing}

The mobile edge computing (MEC) architecture allows content providers to deploy MEC servers at cellular BSs, so that the applications and the tasks processing are performed closer to the cellular users. In a HetNet with SBS equipped with MEC servers, as shown in the {\em 2nd} case of Fig.~\ref{fig4}, the storage and computation capability of MEC stations are limited compared to cloud server. Thus, optimizing the ON-OFF switching should not only consider the traffic load, but also consider the property of each task. For example, a computational intensive task should be directly processed at the cloud server and the nearby SBS should be turned off if it has no other ongoing task. In addition, the processing delay is impacted by both processing speed of MEC server and transmission data rate to an SBS. This brings another factor to be considered in user association, which in turn impacts the BS ON-OFF schedule.

\subsubsection{Wireless Caching Stations}

In a small cell access AP with caching capability, the frequently requested data-intensive tasks can be stored at caching stations in advance, resulting in high QoS for users. As the popular contents vary over time and location, it would be desirable to turn off the under-utilized caching stations for energy saving. Different from cellular BSs, ON-OFF switching of a caching station impacts all the tasks stored in the station. Thus, a thorough consideration is required. The system performance can be further enhanced if the demand of content at each station is predictable, which can be implemented using statistic information or pattern recognition techniques. As shown in the {\em 3rd} case of Fig.~\ref{fig4}, the caching stations can gather the user preference information and forward it to the server in core network. After processing such information with machine learning based approaches, the updated caching contents are sent back to caching stations. The information analysis of user preference can also be executed locally to reduce latency if such capability is available for the caching stations.

\subsubsection{M2M in IoT}

Machine to machine (M2M) communication is expected to be widely applied in the paradigm of Internet of Things (IoT) due to the massive data generated by various types of machines and devices. Similar to D2D, the direct transmission of M2M creates opportunities for APs or BSs to be turned off, as a machine connected to an AP can relay the signals of machines that are out of the coverage. However, this requires additional functionality on the relaying machine/device while most devices in the context of IoT are simple, low-power, low-cost devices act as sensors and not always switched on. Hence, the feasibility of exploiting M2M for energy saving depends on the availability of devices and efficient coordination with low overhead is required.

Although the machines/devices in IoT are only required to connected to the APs occasionally, the upload of the gathered data and the download updated settings must be carried out with a certain frequency due to the storage limit of devices and the required timeliness of data. This requires the APs to be turned on in a proper manner to balance the tradeoff between service provision and energy saving. In the presence of multiple machines with different data size and requirements of update frequency, the task schedule can be jointly considered with AP ON-OFF switching to minimize the long-term energy consumption as well as guarantee the QoS.

\subsubsection{Operation in Unlicensed Band}

When a 5G wireless network operates on unlicensed bands, the coexistence with other systems that already use the spectrum band
is a fundamental challenge. Take WiFi as an example. Due to the differences on media access control (MAC) protocols, WiFi may get stuck in constant backoff states if the 5G network transmits in its normal pattern. For BS ON-OFF switching, a BS may be switched off during certain periods, not only for purpose of energy saving, but also for QoS provisioning for WiFi users. When a BS is turned off, UEs originally served by the BS may handover to the Wi-Fi system for better QoS. Consequently, a tradeoff between 5G UEs and existing system UEs will be an interesting problem to be address in future research.

The Long Term Evolution Unlicensed (LTE-U) is regarded as a promising application scenario in 5G systems. In an LTE-U system, UEs are allocated with orthogonal time-frequency resource blocks (RB). Then, the ON-OFF schedule should be jointly considered with user association and resource allocation of each BS, since these factors determine the aggregated interference level of both LTE and Wi-Fi users. Thus, efficient approaches are required to decouple BS ON-OFF switching with other system configurations. Moreover, since LTE adopts centralized control for service provision with information exchange through X2 interface, the added BS ON-OFF schedule would increase the processing overhead and complexity, it is necessary to design decentralized or computational-efficient schemes to enhance feasibility.

\subsection{Additional Technical Challenges}

\subsubsection{Scalability}

With the expected massive deployment of BSs in 5G networks, the control overhead could be overwhelming. In particular, if software-defined networking architecture is applied, the network control would be performed in a centralized pattern, resulting in prohibitive complexity. To address this, a wireless network may be partitioned into multiple parts to guarantee the feasibility of scheduling. As for BS ON-OFF switching, we can divide BSs into different clusters, each with a controller. At the lower level, intra-cluster scheduling can be performed with reasonable overhead as long as the cluster size is relatively small. At the higher level, different controllers can cooperatively adjust their strategies by taking the inter-cluster impacts into consideration. The cluster size provides a trade-off between complexity and system performance, and the clustering strategy would be an interesting topic for future research.

\subsubsection{Application of Machine Learning Techniques}

With the development of hardware, the machine learning techniques become feasible to network control and can be used in ON-OFF switching to achieve better performance. As we shown in Fig.~\ref{fig1} and Fig.~\ref{fig2}, the tradeoff between energy and delay, and the tradeoff EE and wake up interval are observed. However, if we can predict the traffic pattern, each BS can wake up at proper time instants such that energy saving is achieved without causing additional delay. Then, the machine learning techniques can be applied to predict the user distribution and the possible traffic load of each BS. Based on historic data and the movement of users, the approximate number of users arriving at each BS can be estimated, the expected under-utilized BSs can maintain at sleeping mode. However, the detailed technical aspects to implement this approach still requires further study. Moreover, the learning-based techniques have the potential to be applied with other approaches and yet to be explored in future research.

\subsubsection{QoE Aware Scheduling}

As shown in Table~\ref{tab2}, most existing works consider QoS metrics in system models and problem formulations. However, the actual satisfaction level of a user is determined by multiple QoS factors, such as outage probability, delay, downloading rate, congestion probability, type and format of a video, and energy consumption of mobile devices. Besides, different users have different evaluations over all these aspects. Thus, it is necessary to introduce quality of experience (QoE) in the system design. Moreover, with QoE of users and the energy cost, realistic economic models can be established (e.g. QoE based user payment). The minimization of wireless operator cost can be a new design objective.

\section{Conclusion}

In this article, we aimed to identify the challenges of BS ON-OFF switching in 5G wireless networks and provide insights for potential solutions. We analyzed the technical aspects of BS ON-OFF switching and presented an overview of recent advances in different 5G wireless networks. We concluded this article with a discussion of open problems and outlook.



\end{document}